\documentclass[preprint,aps,nofootinbib]{revtex4}
\usepackage[dvips]{graphicx}
\usepackage{amssymb}
\usepackage{amsmath}
\usepackage{amsmath, amsthm, amssymb, mathrsfs}
\usepackage{pstricks}

\newcommand{\ket}[1]{\lvert #1 \rangle}
\newcommand{\bra}[1]{\langle #1 \rvert}

\begin{document}

\title{Test of Common Sense in Quantum Copying Process}

\author
{
    Mi-Ra Hwang$^1$,
	Eylee Jung$^2$,
	Kap Soo Jang$^3$,
	Mu-Seong Kim$^3$, \\
	DaeKil Park$^{1,3}$,
	Eui-Soon Yim$^4$,
	Hungsoo Kim$^5$,
    Jin-Woo Son$^6$
}
\affiliation
{
    $^1$ Department of Physics, Kyungnam University, Masan, 631-701, Korea \\
	$^2$ Center for Superfunctional Materials,
         Department of Chemistry,
         Pohang University of Science and Technology, 
         San 31, Hyojadong, Namgu, Pohang 790-784, Korea \\
	$^3$ Department of Electronic Engineering, Kyungnam University, Masan, 631-701, Korea \\
    $^4$ Department of Computer Science, Semyung University, Chechon, 390-711, Korea \\
    $^5$ The Institute of Basic Science, Kyungnam University, Masan, 631-701, Korea \\
	$^6$ Department of Mathematics, Kyungnam University, Masan, 631-701, Korea
}

\begin{abstract}
It is believed that the more we have {\it a priori} information on input states, the better we can make the
quality of clones in quantum cloning machines. This common sense idea was confirmed several years ago by analyzing a situation,
where the input state is either one of two non-orthogonal states. If the {\it a priori} information is measured by
the Shannon entropy, common sense predicts that the quality of the clone becomes poorer with increasing $N$,
where $N$ is the number of possible input states. We show, however, that the {\it a priori} information measured
by the Shannon entropy does not affect the quality of the clones. Instead the no-cloning theorem and `denseness' of
the possible input states play important roles in determining the quality. Specifically, the factor `denseness' plays a more
crucial role than the no-cloning theorem when $N \geq 3$.
\end{abstract}


\maketitle

Recently, much attention is being paid to quantum information processing (QIP)\cite{nielsen00}. This is mainly due to
our guess that we may be able to explore beyond classical information processing (CIP) with the aid of QIP. Although
CIP is a base for modern technology, it is well known that it has its own limitations. For example, there is
a time limitation when a classical computer performs a huge numerical calculation. Another example of the limitation of the
CIP is a serious
eavesdropping problem in the classical network. It is believed that such troublesome problems can be overcome if
we substitute QIP for CIP. For this reason the secure quantum cryptography protocols\cite{bb84,ek91} were
developed years ago and are now at the stage of the industrial era\cite{all07,seco09}. Another major application is a quantum
computer, a machine which performs numerical computation within quantum mechanical laws. The quantum computer was
suggested a few decades ago by R. Feynman\cite{feynman1,feynman2}. Currently, much effort has been devoted to the realization
of the quantum computer by making use of various physical setups such as NMR, ion traps, quantum dots, and superconductors.
The current status of the realization is summarized in Ref.\cite{qc10}.

It is needless to say that a perfect copy of a given unknown state is of great help for the real QIP\footnote{Sometimes,
however, a perfect copy of the quantum state can generate some problems in the real QIP. In this case, for example, eavesdropping
can be more easily carried out by an eavesdropper in the cryptographic process.}. However,
the no-cloning theorem\cite{wootters82} forbids a perfect copy of the quantum state.
Nevertheless, it is possible to construct a quantum cloning machine, which produces approximate copies of
the quantum state.
The first trial for the analysis of such a cloning machine was presented by Buzek and Hillery (BH)\cite{buzek96-1} in $1996$.
In modern terminology the BH's cloning machine is a symmetric, state-independent, optimal\footnote{
The term ``symmetric'' means that the quality of the clones are all same.
The term ``state-independent'' means that the quality of clones is independent of the original
input state.
The term ``optimal'' means that the fidelity between original state and clones is maximal.
},
and single qubit $1 \rightarrow 2$ cloning machine, even though BH have not proven the optimality
of their cloning machine in Ref.\cite{buzek96-1}.
Such a cloning machine is usually called a universal cloning machine (UCM).

The optimality of the BH cloning machine was proven
in Ref.\cite{bruss-98-1,zanardi-98,werner-98,keyl-99}.
In addition, authors in Ref.\cite{zanardi-98,werner-98,keyl-99} discussed the optimality
for the cloning of the higher-dimensional quantum state such as multi-qubit or qudit states.
The explicit unitary transformation for the $N \rightarrow M$ UCM is given
in Ref.\cite{gisin-97-1}.

The optimality for the asymmetric cloning machine was also discussed in Ref.\cite{cerf-00,niu-98}.
Especially, in Ref.\cite{cerf-00} Cerf has constructed the Pauli cloning machine (PCM)
on the analogy of the Pauli channel in the decoherence process.
In fact, this is a generalization of the fact that the BH UCM generates the same output state
as that emerging from the depolarizing channel of probability $p=1/4$.
Using PCM, Cerf derived an inequality and guessed that the equal part in this inequality
corresponds to the optimality for the asymmetric cloning machine.
This optimal condition was algebraically proven in Ref.\cite{niu-98} by making use of
the quality measure based on distinguishability rather than the usual fidelity.
However, the optimality for the asymmetric cloning machine seems to depend on the choice of
the quality measure.

The cloning of the higher-dimensional system allows us to research the approximate cloning of entanglement.
Since it is well known that entanglement is a genuine physical
resource\cite{vidal03-1,joz,nielsen00} for the various QIP, this task is important for a practical reason.
In particular, the cloning of the bipartite entanglement such as concurrence and the entanglement
of formation was analyzed in Ref.\cite{masiak-03,novotny-05,karpov-05,choudhary-07-1}.
It has been shown that, in general, much loss of entanglement occurs during the cloning process.
This means that the cloning procedure seems to crucially remove the quantum correlations.
In order to use the cloning machine in the real QIP,
we, therefore, need to find a way to reduce the loss of the entanglement.

In Ref.\cite{bruss-98-1} the optimality for the state-dependent cloning machine
was discussed by making use of a new quality measure called global fidelity.
The remarkable feature of this state-dependent cloning machine is that the cloning procedure
is implemented without an ancilla system. It was shown that, if some {\it a priori} knowledge about the input
state is given, the cloner can perform the cloning procedure much better than the usual universal optimal one.
In fact, this is in agreement with common sense because the more {\it a priori} information one has on a certain experiment,
one can generally tune the experimental setup to produce better outputs. However, the following question naturally
arises: how can we define the {\it a priori} information? In the cloning process the natural choice for the {\it a priori} information
is to measure it in terms of the number of possible input states, that is, the more the number of possible input states there are,
the less the {\it a priori} information we have on the exact input state. Such a measure for the {\it a priori} information is,
in fact, consistent with the
measure defined by the Shannon entropy. Let us consider a set ${\cal S_N}$, which consists of $N$ quantum states. If we
want to copy a single state chosen randomly from the set ${\cal S_N}$, the corresponding Shannon entropy is $\log N$ because
all $N$-states can be an input state with equal probability. Thus, one can say that, if the Shannon entropy increases, the
{\it a priori} information is reduced.

The purpose of this article is to examine
the validity of this common sense idea; that is, {\it a priori} information on the input state can enhance the
quality of the clones. In order to check the validity of this idea we reduce the {\it a priori}
information by increasing the number of possible input states. Under this circumstance we will compute quality measures such
as global fidelity. We will show that, surprisingly, the {\it a priori} information measured by the Shannon entropy does
not affect the quality of the clones. Instead of this, there are two other factors which determine the quality of the
clones.  The first factor is the no-cloning theorem. This factor is very important when the possible input state is one of two known states.
The second factor we have found is the ``denseness'' of the possible input states.
The latter is important when the input state is one of many known states.

Now, we consider the situation that the input state is randomly chosen from a set ${\cal S}_{N}$, which consists of
$N$ states. We first consider the $N = 2n$ case. For simplicity, we choose the $2 n$ states as follows:
\begin{eqnarray}
\label{even-1}
& &\ket{a_m} = \cos \left(m - \frac{1}{2}\right) \theta \ket{0} + \sin \left(m - \frac{1}{2}\right) \theta \ket{1}
    \hspace{1.0cm} (m=1, 2, \cdots, n)                                           \\   \nonumber
& &\ket{b_m} = \cos \left(m - \frac{1}{2}\right) \theta \ket{0} - \sin \left(m - \frac{1}{2}\right) \theta \ket{1}
    \hspace{1.0cm} (m=1, 2, \cdots, n).
\end{eqnarray}	
We also consider a state-dependent cloning machine without an ancilla system, which was introduced in Ref.\cite{bruss-98-1}.
Since $\ket{a_m}$ and $\ket{b_m}$ are on the same plane, the physical role of the copy machine is defined by a
unitary transformation for two of those states, say $\ket{a_n}$ and $\ket{b_n}$. The most general unitary
transformation can be expressed as:
\begin{eqnarray}
\label{even-2}
& &\ket{\alpha_n} \equiv  U \ket{a_n} \ket{ \;\;} = \xi_1 \ket{a_n}\ket{a_n} + \eta_1 \ket{b_n}\ket{b_n} + c_{11} \ket{c_1} + c_{12} \ket{c_2}
                                                                                                                \\   \nonumber
& &\ket{\beta_n} \equiv  U \ket{b_n} \ket{\;\; } = \eta_2 \ket{a_n}\ket{a_n} + \xi_2 \ket{b_n}\ket{b_n} + c_{21} \ket{c_1} + c_{22} \ket{c_2},
\end{eqnarray}
where we have introduced the orthogonal states $\ket{c_1}$ and $\ket{c_2}$, which are also orthogonal to
$\ket{a_n}\ket{a_n}$ and $\ket{b_n}\ket{b_n}$. It is convenient to choose $\ket{c_1}$ and $\ket{c_2}$ as
\begin{eqnarray}
\label{even-3}
& &\ket{c_1} = \frac{1}{\sqrt{\sin^4 \left(n - \frac{1}{2}\right)\theta + \cos^4 \left(n - \frac{1}{2}\right)\theta}}
                                                                                                              \\   \nonumber
& & \hspace{3.0cm} \times
                \left[\sin^2 \left(n - \frac{1}{2}\right)\theta \ket{00} - \cos^2 \left(n - \frac{1}{2}\right)\theta \ket{11} \right]
				                                                                                              \\   \nonumber
& &\ket{c_2} = \frac{1}{\sqrt{2}} \left(\ket{01} - \ket{10} \right).
\end{eqnarray}
Since $\ket{a_m}$ and $\ket{b_m}$ can be written as linear combinations of $\ket{a_n}$ and $\ket{b_n}$, it is straightforward
to compute the unitary transformation for those states. We define those two-qubit states as
\begin{equation}
\label{even-4}
\ket{\alpha_m} = U \ket{a_m} \ket{\;\;}    \hspace{1.5cm}  \ket{\beta_m} = U \ket{b_m} \ket{\;\;}.
\end{equation}
Then the global fidelity for the cloning machine is defined by
\begin{equation}
\label{even-5}
F_g = \frac{1}{2 n} \sum_{m=1}^n \left[ |\bra{a_m}\bra{a_m}\alpha_m \rangle|^2 + |\bra{b_m}\bra{b_m}\beta_m \rangle|^2 \right].
\end{equation}
One can show directly that $\bra{b_m}\bra{b_m}\beta_m \rangle$ can be obtained from $\bra{a_m}\bra{a_m}\alpha_m \rangle$ by
$\xi_1 \leftrightarrow \xi_2$, $\eta_1 \leftrightarrow \eta_2$, and $c_{11} \leftrightarrow c_{21}$. In addition, $F_g$ is
independent of $c_{12}$ and $c_{22}$ due to $\bra{a_m} \bra{a_m} c_2\rangle = \bra{b_m} \bra{b_m} c_2\rangle = 0$.
Since there is no preference in the states $\ket{\alpha_m}$ and $\ket{\beta_m}$, we conjecture that the optimality for the
cloning machine leads to $\xi_1 = \xi_2$, $\eta_1 = \eta_2$, $c_{11} = c_{21}$, and $c_{12} = c_{22} = 0$.
We will show shortly by applying the Lagrange multiplier method that this is indeed the case.

From $\bra{\alpha_n} \alpha_n \rangle = \bra{\beta_n} \beta_n \rangle = 1$ one can derive two constraints
\begin{eqnarray}
\label{even-6}
& &\varphi_1 = |\xi_1|^2 + |\eta_1|^2 + |c_{11}|^2 + |c_{12}|^2 + (\xi_1^* \eta_1 + \xi_1 \eta_1^*) \cos^2 (2n-1)\theta-1
                                                                                                    \\    \nonumber
& &\varphi_2 = |\xi_2|^2 + |\eta_2|^2 + |c_{21}|^2 + |c_{22}|^2 + (\xi_2^* \eta_2 + \xi_2 \eta_2^*) \cos^2 (2n-1)\theta-1.
\end{eqnarray}
Additionally, one can derive the following two constraints from the condition $\bra{\alpha_n}\beta_n\rangle = \bra{a_n}b_n\rangle$;
\begin{eqnarray}
\label{even-7}
& &\varphi_3 = (\xi_1^* \eta_2 + \xi_1 \eta_2^*) + (\xi_2^* \eta_1 + \xi_2 \eta_1^*) + (c_{11}^* c_{21} + c_{11} c_{21}^*)
               + (c_{12}^* c_{22} + c_{12} c_{22}^*)                                               \\    \nonumber
& &  \hspace{1.0cm}
              + \left\{ (\xi_1^* \xi_2 + \xi_1\xi_2^*) + (\eta_1^*\eta_2 + \eta_1 \eta_2^*) \right\} \cos^2 (2n-1) \theta
			  -2 \cos (2n-1) \theta                                                                 \\    \nonumber
& &i \varphi_4 = (\xi_1^* \eta_2 - \xi_1 \eta_2^*) - (\xi_2^* \eta_1 - \xi_2 \eta_1^*) + (c_{11}^* c_{21} - c_{11} c_{21}^*)
               + (c_{12}^* c_{22} - c_{12} c_{22}^*)                                               \\    \nonumber
& &  \hspace{1.0cm}
              + \left\{ (\xi_1^* \xi_2 - \xi_1\xi_2^*) + (\eta_1^*\eta_2 - \eta_1 \eta_2^*) \right\} \cos^2 (2n-1) \theta.
\end{eqnarray}
In order to apply the Lagrange multiplier method for finding the optimal conditions we have to maximize $F_{g}^{\lambda}$ by defining it as
\begin{equation}
\label{even-8}
F_g^{\lambda} = F_g + \sum_{i=1}^{4} \lambda_i \varphi_i.
\end{equation}
Then the optimal conditions are derived from the eight equations $\partial F_g^{\lambda} / \partial \chi = 0$, where
$\chi = \xi_1^*$, $\eta_1^*$, $\xi_2^*$, $\eta_2^*$, $c_{11}^*$, $c_{21}^*$, $c_{12}^*$, and $c_{22}^*$. The remaining
eight equations derived when $\chi = \xi_1$, $\eta_1$, $\xi_2$, $\eta_2$, $c_{11}$, $c_{21}$, $c_{12}$, and $c_{22}$ are
simply complex conjugates to the previous equations. Two equations derived when $\chi=c_{12}^*$ and $c_{22}^*$ are solved by letting
$c_{12} = c_{22} = 0$. This is due to the fact that $F_g$ is independent of $c_{12}$ and $c_{22}$. From the remaining six equations
one can make the following three equations:
\begin{equation}
\label{even-9}
\frac{\partial F_g^{\lambda}}{\partial c_{11}^*} - \frac{\partial F_g^{\lambda}}{\partial c_{21}^*} =
\frac{\partial F_g^{\lambda}}{\partial \xi_1^*} - \frac{\partial F_g^{\lambda}}{\partial \xi_2^*} =
\frac{\partial F_g^{\lambda}}{\partial \eta_1^*} - \frac{\partial F_g^{\lambda}}{\partial \eta_2^*} = 0.
\end{equation}
These equations are solved by imposing $\lambda_1 = \lambda_2$, $\lambda_4 = 0$, $c_{11} = c_{21}$, $\xi_1 = \xi_2$,
and $\eta_1 = \eta_2$. Thus, our conjecture for the optimality is perfectly derived from the Lagrange multiplier method.
In addition to this, $\lambda_4 = 0$ imposes that $\varphi_4$ should be trivially zero. This fact indicates that
all parameters are real. Eventually, we have only two constraints
\begin{eqnarray}
\label{even-10}
& &\tilde{\varphi}_1 = \xi^2 + \eta^2 + c^2 + 2 \xi \eta \cos^2 (2n-1) \theta - 1          \\   \nonumber
& &\tilde{\varphi}_2 = 2 \xi \eta + c^2 + (\xi^2 + \eta^2) \cos^2 (2n-1) \theta - \cos (2n-1) \theta
\end{eqnarray}
and three extrema equations
\begin{equation}
\label{even-11}
\frac{\partial F_g^{\lambda}}{\partial c^*} = \frac{\partial F_g^{\lambda}}{\partial \xi^*} =
\frac{\partial F_g^{\lambda}}{\partial \eta^*} = 0,
\end{equation}
where $\xi_1 = \xi_2 \equiv \xi$, $\eta_1 = \eta_2 \equiv \eta$ and $\lambda_1 = \lambda_2 \equiv \lambda$.
From the constraint equations $\tilde{\varphi}_1 = \tilde{\varphi}_2 = 0$, one can express $\xi$ and $c$ in terms of
$\eta$ as follows:
\begin{eqnarray}
\label{even-12}
& &\xi = \eta + \frac{1}{\sqrt{2} \cos \left(n - \frac{1}{2}\right) \theta}                  \\   \nonumber
& &c^2 = \frac{1}{2} \left[ 1 + \cos (2n-1) \theta -
                             \left\{ 2 \eta + \frac{1}{\sqrt{2} \cos \left(n - \frac{1}{2} \right) \theta} \right\}^2
							 \left[1 + \cos^2 (2n-1) \theta \right]
					 \right].
\end{eqnarray}
Two of the Eq.(\ref{even-11}) are used to determine $\lambda$ and $\lambda_3$ and the remaining one is used to derive an
equation which $\eta$ should obey for optimality. After long and tedious calculations, the following quartic equation is derived:
\begin{equation}
\label{even-13}
z_4 \eta^4 + z_3 \eta^3 + z_2 \eta^2 + z_1 \eta + z_0 = 0,
\end{equation}
where the coefficients have too long expressions to be expressed explicitly. However, those coefficients $z_i$'s can be
straightforwardly derived by manipulating Eq.(\ref{even-11}) and Eq.(\ref{even-12}) appropriately.

For $N = 2n+1$ case we choose the set ${\cal S}_{2n+1}$, which consists of the following states for simplicity:
\begin{equation}
\label{odd-1}
\ket{a_m} = \cos m\theta \ket{0} + \sin m\theta \ket{1}.
\hspace{1.5cm} (m = 0, \pm 1, \cdots, \pm n)
\end{equation}
Since all states are on the same plane, in order to find an optimality it is sufficient to examine the unitary transformation for
$\ket{a_n}$ and $\ket{a_{-n}}$. Let the unitary transformation for those states be
\begin{eqnarray}
\label{odd-2}
& &\ket{\alpha_n} \equiv U \ket{a_n} \ket{\;\;} = \xi_3 \ket{a_n} \ket{a_n} + \eta_3 \ket{a_{-n}} \ket{a_{-n}} +
                            c_{13} \ket{c_3} + c_{14} \ket{c_4}                          \\   \nonumber
& &\ket{\alpha_{-n}} \equiv U \ket{a_{-n}} \ket{\;\;} = \eta_4 \ket{a_n} \ket{a_n} + \xi_4 \ket{a_{-n}} \ket{a_{-n}} +
                            c_{23} \ket{c_3} + c_{24} \ket{c_4}  	
\end{eqnarray}
where $\ket{c_4}$ is identical with $\ket{c_2}$ given in Eq.(\ref{even-3}) and $\ket{c_3}$ is
\begin{equation}
\label{odd-3}
\ket{c_3} = \frac{1}{\sqrt{\sin^4 n\theta + \cos^4 n \theta}} \left[\sin^2 n\theta \ket{00} - \cos^2 n\theta \ket{11} \right].
\end{equation}
Thus, the global fidelity for this case is defined as
\begin{equation}
\label{odd-4}
F_g = \frac{1}{2n+1} \sum_{m=-n}^{n} |\bra{a_m}\bra{a_m}\alpha_m\rangle|^2.
\end{equation}	
Similar analysis to the $N=2n$ case yields $\xi_3 = \xi_4 \equiv \xi$, $\eta_3 = \eta_4 \equiv \eta$, $c_{13} = c_{23} \equiv c$,
and $c_{14} = c_{24} = 0$ for optimality. Also the analysis with Lagrange multiplier method imposes that all parameters are real.
The constraints derived from $\bra{\alpha_n}\alpha_n\rangle = \bra{\alpha_{-n}}\alpha_{-n}\rangle = 1$ and
$\bra{\alpha_{n}}\alpha_{-n}\rangle =	\bra{a_n}a_{-n}\rangle$ with these optimality conditions enable us to express
$\xi$ and $c$ in terms of $\eta$ as follows:
\begin{eqnarray}
\label{odd-5}
& &\xi = \eta + \frac{1}{\sqrt{2} \cos n\theta}                              \\   \nonumber
& &c^2 = \cos^2 n\theta - \frac{1 + \cos^2 2n\theta}{2} \left(2\eta + \frac{1}{\sqrt{2}\cos n\theta} \right)^2.
\end{eqnarray}
By manipulating the extrema equations in similar fashion to the $N=2n$ case one can derive a similar quartic equation
\begin{equation}
\label{odd-6}
\tilde{z}_4 \eta^4 + \tilde{z}_3 \eta^3 + \tilde{z}_2 \eta^2 + \tilde{z}_1 \eta + \tilde{z}_0 = 0.
\end{equation}
Like the previous case the coefficients $\tilde{z}_i$'s are too long to express explicitly. However, the derivation
is straightforward even though it requires complicated calculations. Thus, by solving the quartic equations (\ref{even-13}) and
(\ref{odd-6}), one can compute the $\theta$-dependence of the optimal global fidelity when
the input state is one of the states in ${\cal S}_N$.

\begin{center}
\begin{figure}
    \includegraphics[width = 10 cm]{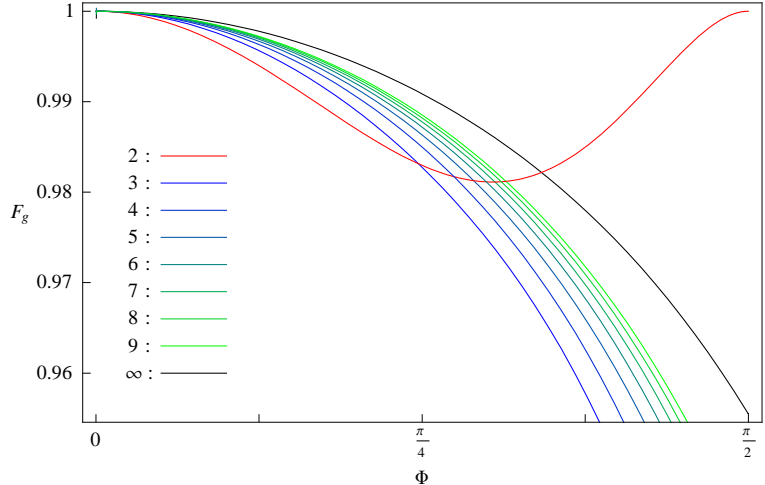}
\caption[fig1]{This figure shows the $\Phi$-dependence of the global fidelity $F_g$ with varying $N$ for the optimal state-dependent
cloning machine. When $N=2$, $F_g$ approaches one at $\Phi \approx 0$ and $\Phi \approx \pi/2$. This reflects on the fact that
the parallel and perpendicular quantum states can be cloned perfectly without violation of the no-cloning theorem. When $N \geq 3$,
$F_g$ with fixed $\Phi$ increases with increasing $N$. This fact is contrary to common sense. This figure strongly suggests that the factor `denseness' of the possible
input states plays an important role in determining the quality of the quantum cloning process.}
\end{figure}
\end{center}

Now, imagine the situation, where all states in ${\cal S}_N$ lie between two vectors. Let the angle between these two
vectors be $\Phi$. Then, we have to rescale $\theta$ as $\Phi / (N-1)$. The $\Phi$-dependence of optimal $F_g$ is plotted
in Fig. 1. The global fidelity for the $N=2$ case (red line) is in strong agreement with Ref.\cite{bruss-98-1}. The global fidelity
for $N=2$ approaches one when $\Phi \approx 0$ and $\Phi \approx \pi/2$. This reflects on the fact that two parallel or
perpendicular states can be perfectly copied without violation of the no-cloning theorem\cite{wootters82}. Thus, the no-cloning theorem
plays a crucial role in determining the quality of the clones when $N=2$. As Fig. 1 has exhibited, the no-cloning theorem
does not play an important role when $N \geq 3$. In these cases the quality measure $F_g$ shows a decreasing behavior with increasing
$\Phi$. However, a surprising result is the fact that $F_g$ with fixed $\Phi$ becomes larger and larger with increasing $N$ and eventually
approaches the limiting value corresponding to $N = \infty$ (black line in Fig. 1).

The limiting value can be computed by considering the continuum case as follows. Let the input state be one of
$\ket{\psi(x)} = \cos x \ket{0} + \sin x \ket{1}$, where $-\Phi/2 \leq x \leq \Phi/2$. Then, one can define a unitary
transformation for two states out of $\ket{\psi(x)}$ in a similar way.
The global fidelity for this continuum case can be defined as a mean value
$$ F_g = \frac{1}{\Phi} \int_{-\Phi/2}^{\Phi/2} dx |\bra{\psi(x)}\bra{\psi(x)} U \ket{\psi(x)}\ket{\;\;}|^2.$$
Then, applying the Lagrange multiplier method to this case and manipulating the constraints and extrema equations
similarly, one can compute the optimal $F_g$ numerically, which is the black line in Fig. 1.

The increasing behavior of the global fidelity with increasing $N$ in fixed $\Phi$ is contrary to common sense
if {\it a priori} information is
measured in terms of the Shannon entropy. Instead, the `denseness'\footnote{the `denseness' can be defined as $N / \Phi$.}
of the states determines the quality of
the cloning output. Although the calculational results of other quality measures are not explicitly presented in this article, 
one can show that they exhibit similar behaviors.
As far as we know, this is a completely new and unknown phenomenon in the cloning process. 
With further research we think that this strange behavior, which
we would like to call the `denseness' effect, can be better understood within quantum mechanical law in the future.

\begin{acknowledgments}
This work was supported by a National Research Foundation of Korea Grant funded by the
Korean Government (2009-0074008).
\end{acknowledgments}

\end{document}